\documentclass[referee, pdflatex,sn-mathphys-ay]{sn-jnl}

\usepackage{caption}
\usepackage{graphicx}%
\usepackage{multirow}%
\usepackage{amsmath,amssymb,amsfonts}%
\usepackage{amsthm}%
\usepackage{mathrsfs}%
\usepackage[title]{appendix}%
\usepackage{xcolor}%
\usepackage{textcomp}%
\usepackage{manyfoot}%
\usepackage{booktabs}%
\usepackage{algorithm}%
\usepackage{algorithmicx}%
\usepackage{algpseudocode}%
\usepackage{listings}%
\usepackage{natbib}
\usepackage{subcaption}
\usepackage{tabularx}
\usepackage{ragged2e}
\usepackage{hyperref}

\theoremstyle{thmstyleone}%
\theoremstyle{thmstyletwo}%

\theoremstyle{thmstylethree}%

\usepackage{tikz}
\usepackage{hyperref}

\begin{document}

\title{Spatial prediction of environmental processes using random forests: How best to account for spatial dependence?}

\author*[1]{\fnm{Duncan}\sur{Lee}(ORCID: https://orcid.org/0000-0002-6175-6800)}\email{Duncan.Lee@glasgow.ac.uk}

\author[1]{\fnm{Vinny} \sur{Davies}(ORCID: https://orcid.org/0000-0003-1896-8936)}\email{Vinny.Davies@glasgow.ac.uk}

\affil[1]{\orgdiv{School of Mathematics \& Statistics}, \orgname{University of Glasgow}, \orgaddress{\city{Glasgow}, \postcode{G12 8SQ}, \country{United Kingdom}}}

\author[2]{\fnm{Helen} \sur{Savage}(ORCID: https://orcid.org/0000-0003-2592-9257)}\email{helenrosecasey@googlemail.com}
\affil[2]{\orgdiv{Department of Microbiology and Infectious Diseases}, \orgname{Bristol NHS Group}, \orgaddress{\city{Bristol}, \country{United Kingdom}}}

\author[3,4,5]{\fnm{Hussein} \sur{Twabi}(ORCID: https://orcid.org/0000-0003-4473-296X)}\email{htwabi@kuhes.ac.mw}
\affil[3]{\orgdiv{Department of Epidemiology}, \orgname{Kamuzu University of Health Sciences}, \orgaddress{\city{Blantyre}, \country{Malawi}}}
\affil[4]{\orgdiv{Chichiri Centre for Health Research and Training}, \orgname{Kamuzu University of Health Sciences}, \orgaddress{\city{Blantyre}, \country{Malawi}}}
\affil[5]{\orgdiv{Institute of Life Course and Medical Sciences}, \orgname{University of Liverpool}, \orgaddress{\city{Liverpool}, \country{UK}}}

\author[6,4]{\fnm{Marriott} \sur{Nliwasa}(ORCID: https://orcid.org/0000-0002-3100-5512)}\email{mnliwasa@gmail.com}
\affil[6]{\orgdiv{Division of Epidemiology and Biostatistics, School of Public Health}, \orgname{University of the Witwatersrand}, \orgaddress{\city{Johannesburg}, \country{South Africa}}}

\author[7]{\fnm{Peter} \sur{MacPherson}(ORCID: https://orcid.org/0000-0002-0329-9613)}\email{Peter.Macpherson@glasgow.ac.uk}
\affil[7]{\orgdiv{School of Health \& Wellbeing}, \orgname{University of Glasgow}, \orgaddress{\city{Glasgow}, \postcode{G12 8TA}, \country{United Kingdom}}}


\abstract{Geostatistical spatial prediction for environmental processes is typically undertaken using Gaussian process models via Kriging, while machine learning (ML) algorithms are state-of-the-art for non-spatial prediction. An exciting recent fusion of these ideas imbibes traditional ML algorithms with the capacity to deal with spatial autocorrelation, leading to improved predictive performance. A range of approaches have been proposed, including fusion with Gaussian processes, observation-driven correlation structures, spatial basis functions and local geographical fitting. However, there has been no numerical comparison of their relative predictive performances, which is needed to advise  environmental scientists on the optimal approach to use. This  paper fills this knowledge gap, and focuses on random forests as the ML algorithm because they are more computationally and conceptually straightforward to implement than deep learning algorithms. The results from two studies are presented, the first being a controlled simulation experiment investigating whether any single approach is consistently superior across different spatial autocorrelation types.  The second study focuses on the prediction of air pollution concentrations within a tuberculosis prevalence study in Blantyre, Malawi. The results show that whilst no single approach is universally superior, utilising spatial basis functions appears to perform consistently well across both the simulation and real data studies.}


\maketitle

\section{Introduction}

Environmental processes such as concentrations of air pollution are typically only measured at a small number of locations, due to both the availability and cost of the measuring devices and the accompanying expertise required to make them operational. However, users of these data require measurements on a dense grid of points across a wide region of interest, which can hence be visualised on a map. For example, in the context of air pollution that forms the motivating study in this paper, a national environment agency needs to assess whether concentrations across their country comply with global regulatory standards such as the World Health Organization air quality guidelines (\citealp{who2021guidelines}), while epidemiological researchers investigating the health impacts of air pollution need predicted exposures for every member of their study population (e.g., \citealp{WHO2025}). As a result, spatial prediction of environmental processes at locations where it has not been measured is a vital task to fully unlock the power of these data, and hence it has been the subject of much methodological development.

The standard approach to spatial prediction for point-level data is via Kriging (\citealp{krige1951}) or its Bayesian implementation (\citealp{Banerjee2014}), because it is the best linear unbiased prediction in a mean square error sense. Numerous methodological advancements have been based on this framework, including addressing the spatial change of support problem (\citealp{Gelfand2001}), fusing measured and modelled data products into a single prediction model (\citealp{Berrocal2010}), and incorporating a preferential sampling mechanism to better reflect how the measurement locations were chosen (\citealp{diggle2010}). Non-spatial machine learning (ML) algorithms such as random forests (\citealp{breiman2001}) and neural networks  (\citealp{lecun1990}) have also been used to predict environmental processes, and a comparison of different machine learning and Kriging-based approaches in an air pollution context was undertaken by \cite{berrocal2020}. An exciting avenue of current research fuses machine learning algorithms with spatial prediction models such as Kriging, and recent narrative reviews of the literature are given by \cite{wikle2023} and \cite{Patelli2024}. The power of these fusion approaches lies in the fact that the  machine learning and Kriging paradigms use different aspects of the spatial data to make predictions, with the former utilising  complex nonlinear covariate-response relationships whilst ignoring spatial autocorrelation, while the latter models this autocorrelation at the expense of simpler and mainly linear pre-specified covariate-response relationships.

The simplest such fusion approaches capture spatial autocorrelation within the machine learning algorithm by augmenting the covariate set with additional spatially smooth quantities, such as spatially smoothed covariates / responses (\citealp{soltani2022}),  spatial basis functions (\citealp{chen2024}), or by spatially smoothing the nodes in a neural network model (\citealp{zhu2022}). Alternatively, \cite{georganos2021} propose a spatially local fitting procedure for each prediction location, which extends geographically weighted regression to a random forest context. Finally, a spatial autocorrelation model can be incorporated into an extended ML algorithm, by either modifying the objective function (\citealp{saha2023} and \citealp{Zhan2025}), applying a de-correlation pre-process step to the data (\citealp{heaton2024}), or by sequentially fitting ML and spatial autocorrelation models in a residual learning stacked modelling strategy (\citealp{hengl2015} and \citealp{macbride2025}). All of these spatial ML approaches show superior predictive performance compared to a standard non-spatial ML implementation, but the currently unanswered question is does one of these approaches consistently outperform the others, or do their relative performances depend on the cause of the spatial autocorrelation present in the data?

The main aim of this paper is to address this important unanswered question, and hence provide environmental scientists and other researchers with guidance on how best to account for spatial autocorrelation within a machine learning prediction algorithm. We note that purely narrative reviews have been provided by \cite{wikle2023} and \cite{Patelli2024}, but here we proffer the first numerical comparison of their relative predictive performances across a range of real and simulated data sets. As the number of potential ML algorithms is large we restrict attention to just one, because it allows for a focused assessment of how best to allow for spatial autocorrelation within this class of ML algorithms without the results being  obfuscated by a comparison across multiple ML methods. Random forests are the prediction algorithm we utilise in this paper for two reasons. Firstly, they are much more computationally efficient and straightforward (less tuning parameters to consider) to implement than other algorithms such as deep learning neural networks, which hence makes them more accessible to researchers outside the statistics / machine learning community. Secondly, existing research shows that they often outperform neural networks for tabular (\citealp{Grinsztajn2022}) and spatial (\citealp{macbride2025}) data, which is the type of data considered in the motivating study. We provide a review of the set of spatially adapted random forest approaches proposed in the literature in Section 3, which also includes their extension to the current context in some cases. Then a formal comparison of their relative predictive performances is presented in Sections 4 and 5 in the form of two separate studies. The first is a comprehensive simulation experiment (Section 4) comparing their relative predictive performances across a range of spatial autocorrelation types. The second (Section 5) is an out-of-sample prediction experiment applied to an epidemiological study based in Blantyre, Malawi, where predictions of air pollution are required to assess its impact on tuberculosis infection and disease burden. The background and motivation for this application is presented in Section 2, while the paper concludes in Section 6 with a summary and recommendations for how best to model spatial dependence when using a random forest algorithm.

\section{Air pollution prediction for the SCALE study}

\subsection{Setting and population}
Blantyre City is located in the Southern Region of Malawi, and was home to 800,264 people in 2018. It contains a mixture of urban and peri-urban densely-populated informal settlements with poor access to health services, and more established neighbourhoods, which are interspersed with mountainous terrain. With population growth there has been extensive deforestation of mountain areas and an expansion of highways, resulting in a worsening air quality indicators (\citealp{lim2024}).

The Sustainable Community-based Active case finding for Lung hEalth (SCALE) trial was conducted in Blantyre between 30$th$ April 2019 and 13$th$ March 2020 to investigate the effectiveness of community-based active case finding for tuberculosis (TB). Methods and results of the SCALE trial have been previously reported (\citealp{feasey2023}), but in brief, prior to trial interventions being implemented, a household prevalence survey for TB and HIV was conducted in 72 neighbourhood clusters purposively sampled for having a high burden of disease. These clusters are displayed in Figure \ref{fig:spacetimepollution} (grey lines), and  were defined based on community health worker catchment areas. Based on a 2015 citywide census it is estimated that approximately three-quarters of Blantyre City's population are located within these clusters. 

The SCALE study randomly sampled 115 households from each cluster, which were visited by researchers who undertook a survey recording household and dwelling characteristics using electronic questionnaires. Each household was only surveyed once, but was visited a maximum of three times between 8am and 4pm on all days of the week to attempt to complete the survey. The researchers rotated through the 72 clusters in sequence, aiming to complete the recruitment of two clusters in each week. This resulted in $N=7,173$ households being surveyed in total across these 72 spatial clusters, with household locations being the unit of inference in this study. The study design means that each household was sampled on only one day, and for convenience  nearby clusters were typically sampled close together in time. The exact temporal patterning of the sampling across the 72 clusters is summarised in Section S1.1 of the supporting material.

\subsection{Air pollution measurements}
Approximately half of the researchers were issued with Purple Air Classic (PurpleAir Inc, Bluffdale, USA) air quality monitors, initially resulting in $3,363$ households having indoor air pollution measurements while the remaining  $3,810$ households did not have any  measurements. Purple Air Classic monitors have dual laser particle counters (sensor A and sensor B) that measured  concentrations of particulate matter with an aerodynamic diameters less than or equal to 2.5$\mu m$ (PM$_{2.5}$). Following the manufacturer's instructions, the means of particle sensors A and B were taken in each time period that the monitor was operational for in each household. The set of households have different numbers of measurements, with a median of 7 and 1$st$, 25$th$, 75$th$ and 99$th$ percentiles of 1, 4, 10 and 55 respectively. Exploratory analyses showed that a small proportion of these measurements are implausible and likely to be instrument error, such as in one household where two PM$_{2.5}$ measurements of 481.0$\mu gm^{-3}$ and 36.3$\mu gm^{-3}$ were recorded only 4 minutes apart. Therefore, we summarise the concentrations at each household using robust statistics, including the median as a best estimate of the average concentration  and the median absolute deviation from the median (MAD) as a measure of within household variation. 

As some of these household-level measurements are highly likely to be errors (see example above) we remove those observations that meet either of the following two criteria. The first is that the household only has one PM$_{2.5}$ measurement that hence cannot be corroborated. This results in 110 measurements being removed, and we  empirically observed that a sizable proportion of the implausibly large PM$_{2.5}$ concentrations in our data set were singleton measurements.  The second exclusion criteria was when the within-household variation in the measurements was too high to be plausible, which we measure using the robust MAD. Choosing a cutoff for the MAD is obviously arbitrary, but looking at its distribution showed that values above 30 empirically appeared to be outliers and were hence removed (30 in total). This data cleaning resulted in $N_1=3,223$ households with PM$_{2.5}$ concentrations, while $N_2=3,950$ households did not have such data and hence require predictions to be made. 

\begin{figure}[p]
    \centering
    \begin{subfigure}{\textwidth}
        \centering
        \includegraphics[width=0.8\linewidth]{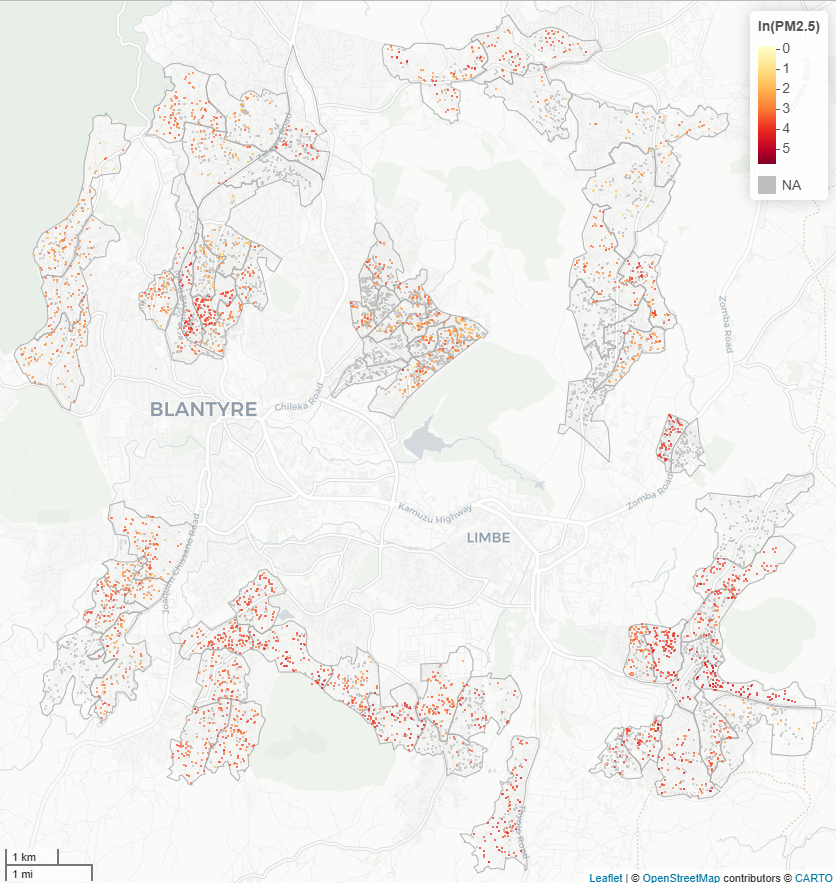}
        \caption{Spatial trend in $\ln(\mbox{PM}_{2.5})$.}
    \end{subfigure}
    \vfill 
    \begin{subfigure}{\textwidth}
        \centering
        \includegraphics[width=0.8\linewidth]{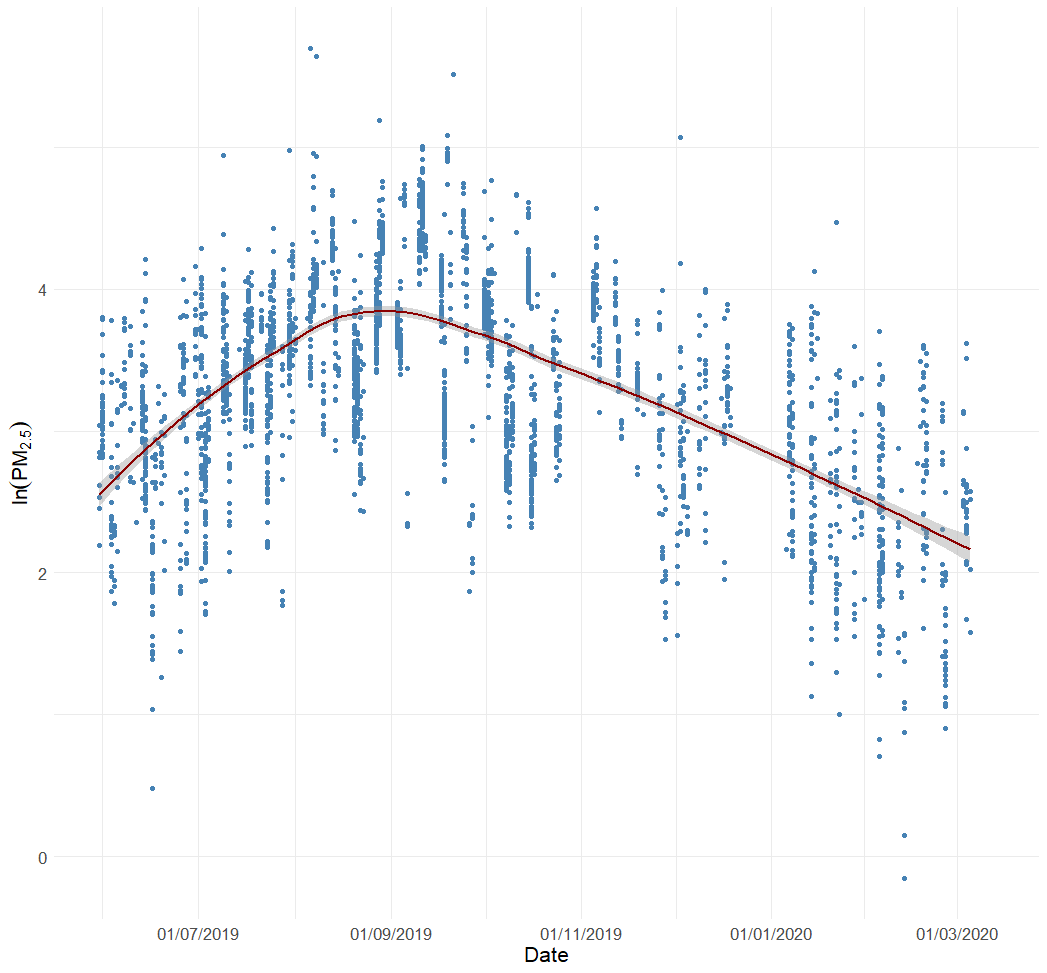}
        \caption{Temporal trend in $\ln(\mbox{PM}_{2.5})$.}
    \end{subfigure}

    \caption{Summary of the spatial (top) and temporal (bottom) trends in the median household level $\ln(\mbox{PM}_{2.5})$ concentrations. In panel (a) the grey dots relate to households with no pollution measurements where predictions are required. In panel (b) the red line is a simple LOWESS smooth.}
    \label{fig:spacetimepollution}
\end{figure}

A histogram of these household-level median concentrations is displayed in Section S1.2 of the supporting material, which shows that they are substantially skewed to the right. Initial analyses additionally showed that this skewness  is retained in all model residuals, so all modelling in this paper is done on the natural log scale, with the resulting inferences being back-transformed to the original scale. The spatial and temporal trends in the log median concentrations are displayed in Figure \ref{fig:spacetimepollution}, which shows that the highest concentrations were recorded in the south and east of the region during September 2019. However as outlined above, the sampling design makes it impossible to clearly attribute the trends shown solely to either space or time.

\subsection{Covariates}
The SCALE survey collected data on a number of potential covariates that might influence air pollution concentrations, which are listed in Section S1.3 in the supporting material. In brief, the first set of covariates relate to the household and dwelling size, including the number of people and the number of rooms in the dwelling. The second set relate to the power sources used in the dwelling, including the main fuel sources for cooking and lighting, location within the dwelling of cooking and whether or not the household was connected to the electricity grid. The main fuel sources used for cooking and lighting had 27 and 100 missing observations respectively, which were imputed using the 5 nearest neighbours algorithm based on the remaining covariates. 

The next set of covariates relate to socio-economic deprivation, including a 5-level education variable, the number of people in the household who receive a regular salary, whether the household have worried about not having enough food in the last week, and a 6-level poverty variable. The latter was a result of asking the head of the household to rank their household position on a series of steps ranging from 1 (poorest people) to 6 (richest people), with this approach having previously been validated in the Malawi national Integrated Household Survey (see \url{https://microdata.worldbank.org/index.php/catalog/2936}). Meteorological measurements (temperature, humidity, pressure) were also made by the air quality monitors, with values at the remaining households being imputed by computing the mean value at households where it has been measured on the same (or closest) day (meteorology should be similar in spatially neighbouring clusters on  the same day). The next set of covariates were generated from publicly available gridded data sources, including the distance to the nearest road in kilometres, (log) population density, and the density of buildings in the surrounding area. Finally, we also use the day and time when each household was surveyed, which are summarised by hour of the day, day of the week (categorical) and the month of the study (categorical) variables.

\subsection{Study aim}
The main aim of analysing these data is to predict PM$_{2.5}$ concentrations at the $N_2=3,950$ households that did not have any (or reliable) measurements, and these predictions will be used in a future epidemiological study examining the impact of elevated PM$_{2.5}$ concentrations on TB infection and disease burden. A secondary aim is to examine which of the covariates have the greatest impact on PM$_{2.5}$ concentrations, and the analysis addressing both of these aims is presented in Section 5.

\section{Methodology}
This section reviews and in some cases extends the classes of methodological approaches proposed in the literature for adapting random forest algorithms to predict geostatistical spatial data, all of which attempt to capture the inherent spatial smoothness  that is ubiquitous in such data. We consider a set of $i=1,\ldots,N$ locations $\{\mathbf{s}_i=(s_{1i}, s_{2i})\}_{i=1}^{N}$, where the first $i=1,\ldots,N_1$ locations have measured observations while the remaining $i=N_1+1,\ldots,N$ locations do not have measurements and hence require predictions to be made. The $N\times 1$ vector of random variables representing the response variable to be predicted is denoted by $\mathbf{Y}=[Y(\mathbf{s}_1),\ldots,Y(\mathbf{s}_N)]^{\top}$, while the corresponding $N\times p$ matrix of covariates are denoted by $\mathbf{X}=[\mathbf{x}(\mathbf{s}_1),\ldots,\mathbf{x}(\mathbf{s}_N)]^{\top}$, where the $p\times 1$ vector $\mathbf{x}(\mathbf{s}_i)=[x_1(\mathbf{s}_i),\ldots,x_p(\mathbf{s}_i)]^{\top}$ denotes the covariates for location $\mathbf{s}_i$. We begin by reviewing the standard random forest algorithm in Section 3.1, before outlining the range of methodological adaptations proposed in the literature for incorporating spatial smoothness in Sections 3.2 to 3.5. 

\subsection{Random forest}
Random forests (RF) were developed by \cite{breiman2001}, and are one of the most popular and best performing machine learning prediction algorithms (\citealp{boehmke2020}). They represent the response variable as

\begin{eqnarray}
    Y(\mathbf{s}_{i})&=& m[\mathbf{x}(\mathbf{s}_i)] + \epsilon(\mathbf{s}_i)\hspace{1cm}\mbox{for }i=1,\ldots,N,
\end{eqnarray}

which is an additive decomposition of the observed data into the true values $\{m[\mathbf{x}(\mathbf{s}_i)]\}$ and random errors $\{\epsilon(\mathbf{s}_i)\}$. The latter are  assumed to be independent and identically distributed with some distribution $g(.)$ across both the observation and  prediction locations. Random forests estimate $\{m[\mathbf{x}(\mathbf{s}_i)]\}$ with an ensemble of  $N_{tr}$ regression trees, where each tree is used to predict all data points before averaging these predictions over all $N_{tr}$ trees to yield the final predictions. Here we fix $N_{tr}=1,000$ (simulation study) and $N_{tr}=2,000$ (Malawi study), which initial analyses in both contexts showed was sufficient for the prediction error to stabilise. Each regression tree is fitted to an independent bootstrapped sample of the data of the same size (i.e., $N_1$), where the sampling is done independently with replacement. Given a bootstrapped data sample, a tree is built using the recursive binary partitioning algorithm of \cite{breiman1984}, which considers a random subset of \texttt{mtry} covariates when making each split in the tree. This recursive splitting continues while all nodes in the tree contain at least \texttt{minnode} observations. In this paper we  consider candidate values for \texttt{minnode} of $\{5, 10, 20\}$ representing different tree complexities, while for \texttt{mtry} we consider values that correspond to $\{5\%, 10\%, 20\%, 30\%, 50\%\}$ of the total number of covariates. 

The optimal values of these tuning parameters are chosen by minimizing the out-of-sample root mean square prediction error (RMSPE), which is approximated using the out-of-bag (OOB) predictions obtained from the random forest for each data point. Specifically, as each regression tree is fitted to a bootstrapped (with replacement) data set, only around 63\% of the distinct data points appear in each bootstrap sample. This means that approximate out-of-sample predictions can be made for each data point by averaging the predictions from the sub-set of trees fitted without the data point in question. These are known as out-of-bag predictions. Finally, as random forests only provide a single point prediction without a measure of predictive uncertainty, we utilise the quantile random forest approach for producing 95\% prediction intervals proposed by \cite{meinshausen2006}.

\subsection{Spatially smoothed features}
The simplest approach is to incorporate spatially smoothed continuous covariates and / or responses as additional covariates in the random forest, which was proposed by \cite{soltani2022} and is akin to the ideas used in spatial autoregressive models (see \citealp{haining2021}) and graphical convolutional neural networks  (see \citealp{zhu2022}). These approaches directly account for  spatial spillover effects  where the environmental process or its continuous risk factors at a given location directly influence its value at nearby locations, for a discussion see \cite{haining2021}. Consider a generic continuous variable $[z(\mathbf{s}_1),\ldots, z(\mathbf{s}_{N_1})]$ observed at the $N_1$ data locations, which could be one of the $p$ covariates or the response. Then a spatially smoothed version of this variable denoted by $\tilde{\mathbf{z}}=[\tilde{z}(\mathbf{s}_1),\ldots, \tilde{z}(\mathbf{s}_N)]$ can be computed across all $N$ observation and prediction locations as

\begin{equation}
\tilde{z}(\mathbf{s}_i)=\frac{\sum_{j=1}^{N_1}w_{ij}z(\mathbf{s}_j)}{\sum_{j=1}^{N_1}w_{ij}}~~~~\mbox{and}~~~~w_{ij}=\exp(-\gamma d_{ij})~~~~\mbox{for }i=1,\ldots,N,  \label{eq_weighedav}  
\end{equation}

where $w_{ii}=0$ to ensure that a data point does not influence its own smoothed value. The smoothed covariate is a weighted average of the observed data, where the weights $\{w_{ij}\}$ are non-negative decreasing functions of the Euclidean distance $\{d_{ij}=||\mathbf{s}_i - \mathbf{s}_j||\}$ between the two locations in question, so that the smoothed variable represents the value in its local neighbourhood. We use an exponential kernel for the weights because it aligns with the exponential autocorrelation function commonly used in geostatistics, as well as its desirable property that the the kernel is bounded between $[0,1]$ ensuring that the weights do not blow up for very small distances as an inverse distance kernel would. The value of gamma controls the distance-decay rate of the weights, with larger values leading to the spatially smoothed covariate being more focused on the values in its immediate vicinity. We treat $\gamma$ as an additional tuning parameter in the algorithm, which is optimised by minimising the out-of-bag root mean square prediction error as before. We consider 5 candidate values of $\gamma$  which correspond to the closest 5\% of locations accounting for 30\%, 50\%, 70\% and 90\% of the total weight (averaged over all data points) in (\ref{eq_weighedav}).

\subsection{Spatially smooth basis functions}
Spatially smooth basis functions can be used as covariates within a random forest, which has the advantage over the previous approach of being implementable regardless of how much covariate information is available in the data application. A number of different types of spatial basis functions have been proposed, and here we utilise the multi-resolution radial basis functions constructed with a Wendland compactly supported correlation function proposed by \cite{nychka2015}. These basis functions were recently incorporated into a simple feed forward neural network by \cite{chen2024}, so here we adapt that approach by utilising them within a random forest. Starting at the most coarse resolution, a rectangular grid of $Q$ equi-spaced knots $\{\mathbf{u}_1,\ldots,\mathbf{u}_Q\}$ is specified, and a basis function is centered at each knot. The value of the basis function centered at $\mathbf{u}_j$ and evaluated at $\mathbf{s}_i$ is given by $\phi(\mathbf{s}_i; \mathbf{u}_j)= \phi^{*}(d=||\mathbf{s}_i-\mathbf{u}_j|| / \theta)$, where

\begin{equation}
\phi^{*}(d)=\left\{\begin{array}{ll}\frac{(1-d)^{6}(35d^2 + 18d + 3)}{3} &\mbox{if~} d\in[0,1]\\
    0 & \mbox{Otherwise}\end{array}\right..
\end{equation}

Here, $\theta$ is a scale parameter controlling the spatial compactness of the basis functions, and we set $\theta$ equal to 2.5 times the between knot distances following the recommendation of \cite{nychka2015}.  The knots at each subsequent higher resolution are twice as dense as those at the previous resolution, resulting in vastly increasing numbers of basis functions as the resolution increases. In this application we consider up to 3 resolutions, where the knots at the most coarse resolution are 5$km$ apart. This results in 25, 81 and 289 basis functions in each of the 3 resolutions, and the maximum number of resolutions (either, 1, 2 or 3) incorporated into the model is treated as a tuning parameter and optimised by minimising the out-of-bag root mean square error as above. 

\subsection{Local data fitting}
An alternative use of the spatial information in the data is to predict the value at a given location using a  \emph{local} random forest that only uses nearby, and hence likely similar, data points in the model fitting. This geographical random forest (GRF) approach is an extension of geographically weighted regression (\citealp{fotheringham2003}) to  a random forest context, and was proposed by \cite{georganos2021}. It predicts the response variable $Y(\mathbf{s}_i)$  by

\begin{equation}
  \hat{Y}(\mathbf{s}_i)  ~=~\alpha \hat{Y}^{local}(\mathbf{s}_i) + (1-\alpha)\hat{Y}^{global}(\mathbf{s}_i),
\end{equation}

where $\{\hat{Y}^{global}(\mathbf{s}_i), \hat{Y}^{local}(\mathbf{s}_i)\}$ denote predictions from global (i.e., a standard) and  local random forest models. Here, $\alpha$ is a weight parameter in the interval $[0,1]$, with larger values giving more weight to the prediction from the local model. This local model is fitted to data from the nearest $b_w$ locations with observed values, which along with $\alpha$ are the two additional tuning parameters of the algorithm. Optimised values are chosen using the out-of-bag approach outlined above, where the candidate sets are $\alpha\in\{0.25, 0.5, 0.75, 1\}$ ($\alpha=0$ corresponds to a standard random forest) and $b_w\in\{100, 500\}$.
 Unlike the initial GRF implementation by \cite{georganos2021}, we construct 95\% prediction intervals based on a normal approximation and the out-of-bag variance estimation approach following the work of \cite{zhang2020}. Note, we do not utilise the quantile regression forest approach here because it is not straightforward to combine prediction intervals from a local and a global random forest. 
 
\subsection{Spatially smooth Gaussian process} 
The final approach we consider combines a random forest with a spatially smooth Gaussian process. We combine these models using residual learning via a stacked modeling strategy, which sequentially applies the random forest and Gaussian process models to the data by using the output from one model as an input to the next model. The simplest such approach proposed by \cite{hengl2015} applied the Gaussian process model to the residuals from a random forest, before combining the predictions from both model components to get the final predictions. More recently, \cite{macbride2025} and \cite{figueira2025} extended this approach by proposing iterative algorithms that sequentially re-fit the random forest and spatial smoothing models to the data, where the predictions from one model component are incorporated into the other model. 

Here, we compare both one-step and full iterative approaches, where in both cases the random forest is fit to the data first and its tuning parameters are optimised by minimising the OOB RMSE as before. Out-of-sample predictions are then made from the tuned RF at both observation (using OOB predictions) and prediction locations, which are then incorporated into the Gaussian process model as a fixed offset. The Gaussian process model is fitted in a Bayesian setting using the integrated nested Laplace approximation stochastic partial differential equation (INLA-SPDE) approach proposed by \cite{Lindgren2011}, which jointly models the data at observed and prediction locations where the latter are treated as missing observations. This second stage model has the general form

\begin{equation}
    Y(\mathbf{s}_i)=\beta_0 + \hat{m}[\mathbf{x}(\mathbf{s}_i)] + \phi(\mathbf{s}_i) + \epsilon(\mathbf{s}_i),~~~~\epsilon(\mathbf{s}_i)\sim\mbox{N}(0, \sigma^2)~~~~\mbox{for }i=1,\ldots,N,\label{eq_GP}
\end{equation}

where $\{\hat{m}[\mathbf{x}(\mathbf{s}_i)]\}$ are the predictions from the random forest. The Gaussian process evaluated at all $N$ observation and prediction locations is modelled by $\boldsymbol{\phi} =[\phi(\mathbf{s}_1),\ldots,\phi(\mathbf{s}_N)]\sim\mbox{N}(\mathbf{0}, \boldsymbol{\Sigma}(\boldsymbol{\theta}))$, where the spatial covariance matrix $\boldsymbol{\Sigma}(\boldsymbol{\theta})$ is represented by a Mat\'{e}rn autocovariance function. We fix the smoothness parameter $\nu=1$ so that the process is mean square differentiable once, which allows a sensible compromise between the exponential ($\nu=0.5$) and Gaussian ($\nu=\infty$) models and is the default recommended by the INLA software. Penalised complexity priors are specified for the spatial range and partial sill parameters as suggested by the INLA software, and initial analyses showed these had little impact on model performance. The SPDE implementation of the Gaussian process is based on a Gaussian Markov random field approximation  fitted on a mesh of triangles. Initial analyses investigated a number of different mesh configurations, and a relatively dense mesh of 27,093 triangles was chosen because increasing this number did not improve the out-of-sample predictive RMSPE. 

For the full iterative algorithm the posterior mean of the spatial process $\boldsymbol{\phi}$ is then subtracted from the response variable when fitting the next iteration of the random forest, which thus allows feedback and evolution between the two model components. This two-step process is repeated for a maximum of 10 iterations, with the optimal number of iterations chosen by minimising the widely applicable information criterion (WAIC, \citealp{watanabe2010})  because it is asymptotically equivalent to the Bayes predictive cross-validation loss. Final inference for both the one-step and iterative algorithms is based on 1,000 draws from the posterior predictive distributions from the Gaussian process model, which are summarised by the posterior mean and 95\% prediction intervals.  An algorithmic formulation of the full iterative approach is given in Section S2 of the supporting material, and is based on the algorithm of \cite{macbride2025}

\section{Simulation study}
This section presents a simulation experiment, whose aim is to determine whether one approach to accounting for spatial autocorrelation when making predictions within a random forest always outperforms the others, or whether the optimal approach depends on the cause of the spatial autocorrelation in the data. The study design is outlined in Section 4.1 while the results are presented in Section 4.2.

\subsection{Study design}
Each simulated data set is generated at the $N_1=3,223$ locations where PM$_{2.5}$ concentrations were measured in the SCALE study, which are illustrated in Figure \ref{fig:spacetimepollution}. A random sample of 80\% of these data points are used to fit each model, while the remaining 20\% of the locations are used to assess out-of-sample predictive performance. Each  simulated data set consists of a continuous response $\{Y(\mathbf{s}_i)\}_{i=1}^{N_1}$,  eight covariates $\{\mathbf{x}(\mathbf{s}_i)=[x_1(\mathbf{s}_i),\ldots,x_{8}(\mathbf{s}_i)]\}_{i=1}^{N}$ and the easting and northing spatial coordinates of the household locations. The first four covariates are generated independently in space from standard normal distributions, which are subsequently rescaled to the unit interval. The remaining four are generated as spatially autocorrelated from a zero-mean Gaussian process with an exponential autocorrelation matrix. The range parameter controlling the strength of this autocorrelation is fixed so that the autocorrelation is 0.75 for locations at the 5$th$ percentile of the set of pairwise distances between locations to ensure relatively strong autocorrelation. These covariates are also rescaled to the unit interval.

The response variable is generated as $Y(\mathbf{s}_i)=\mu(\mathbf{s}_i) + \epsilon(\mathbf{s}_i)$, where $\mu(\mathbf{s}_i)$ denotes the true value and $\epsilon(\mathbf{s}_i)$ introduces zero-mean Gaussian measurement error with variance $\sigma^2$. The true value is modelled as $\mu(\mathbf{s}_i)=m[x(\mathbf{s}_i)] + C_i$, where $m[\mathbf{x}(\mathbf{s}_i)]$ is a non-linear function of the covariates, while $C_i$ corresponds to additional spatial autocorrelation in the true values that does not result directly from the covariate component. The covariates component is given by $m[\mathbf{x}(\mathbf{s}_i)]= 3 / \{1 + \exp[6 - 12x_1(\mathbf{s}_i)]\} + 2[x_3(\mathbf{s}_i)x_6(\mathbf{s}_i)-0.4]^2 +  \cos[2\pi x_7(\mathbf{s}_i)] + x_8(\mathbf{s}_i)$, which includes linear and non-linear relationships, interactions between covariates, as well as some covariates that have no effect on the response. The covariate and correlation components are subsequently scaled so that they respectively account for $2/3$ ($m[\mathbf{x}(\mathbf{s}_i)]$) and $1/3$ ($C_i$) of the spatial variation in the true value of the response. We consider 6 different scenarios in this study, which are all pairwise combinations of the following two factors. The first is the level of measurement error (noise) in the response, and we evaluate relative model performances in low and high noise scenarios by setting $\sigma^2=0.1^2$ and $\sigma^2=1^2$. The second is the cause of the spatial autocorrelation, and we consider the following 3 cases.

\begin{itemize}
    \item \textbf{1. Unmeasured confounding} - Not having data on an important spatially smooth risk factor is a common cause of residual spatial autocorrelation. We mimic this by representing $\{C_i\}$ with a covariate generated from a zero-mean Gaussian process with a spherical autocorrelation structure that decays to zero at the 50th percentile of pairwise distances between all locations. This covariate influences the response but is unobserved, and hence is not included in any models.  
    
    \item \textbf{2. Spatial spillover effects} - Covariates are likely to affect the response at neighbouring locations as well as at their own location, which is another common cause of residual spatial autocorrelation. We mimic this spatial spillover by representing $\{C_i\}$ as the sum of spatially weighted versions of two of the covariates $\{x_1(\mathbf{s}_i), x_6(\mathbf{s}_i)\}$, where the spatial weights are generated as outlined in (\ref{eq_weighedav}). 

    \item \textbf{Spatially-heterogeneous effects} - The covariate-response relationships can often vary over space due to unobserved local factors, which thus induce spatial autocorrelation into the residuals obtained from models that estimate single global effects. We mimic this by splitting the region into five sub-regions, and generating data with different covariate response relationships for each sub-region. Note, here $\mu(\mathbf{s}_i)$ is not partitioned into $\{m[\mathbf{x}(\mathbf{s}_i)], C_i\}$ as in the other scenarios, but instead  $m[\mathbf{x}(\mathbf{s}_i)]$ is varied over the five sub-regions.
\end{itemize}

Five different random forest frameworks are compared in this study, beginning with a standard non-spatial implementation (denoted \texttt{RF}). The remaining four RF frameworks allow for spatial smoothness in the data by: (i) incorporating spatially smoothed features as additional covariates (Section 3.2); (ii) incorporating spatial basis functions as additional covariates (denoted \texttt{RF-Basis}, Section 3.3); (iii) using local data fitting (denoted \texttt{RF-Local}, Section 3.4); and (iv) combining an RF with a Gaussian process model in one-step (denoted \texttt{RF-GP1}) and full iterative (denoted \texttt{RF-GPfull}) algorithms (Section 3.5). For (ii), we implement three related approaches that incorporate the following spatially smoothed components as additional covariates: (a) \texttt{RF-SmoothX} - covariates; (b) \texttt{RF-SmoothY} - response; and (c) \texttt{RF-SmoothXY} - covariates and response.

\subsection{Results}
Fifty simulated data sets are generated under each of the 6 scenarios, and model performance is summarised below (point prediction and speed) and in Section S3.2 of the supporting material (uncertainty quantification).

\subsubsection{Accuracy of point prediction}
\begin{figure}
    \centering
   \includegraphics[width=0.9\linewidth]{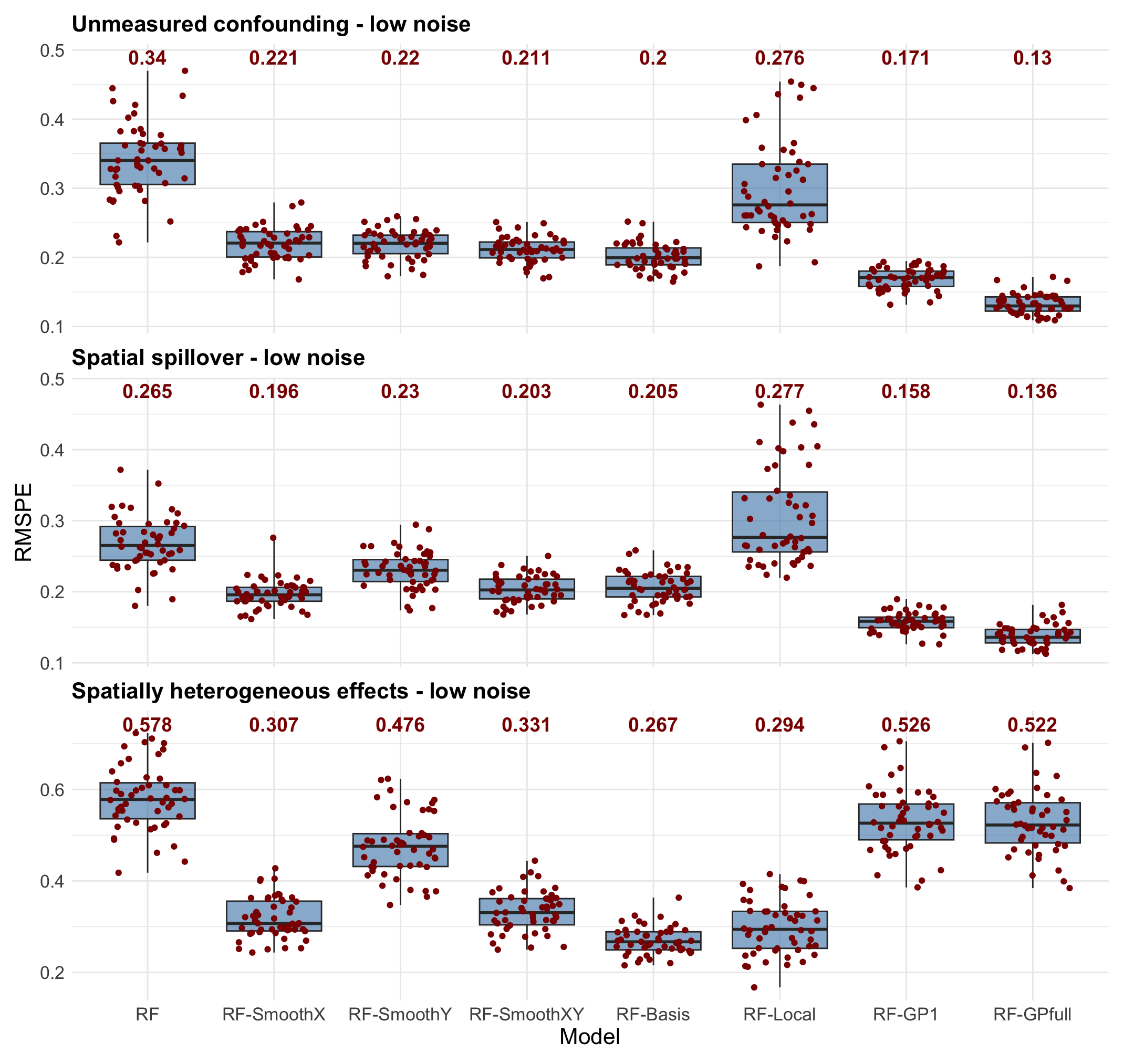}
    \caption{Boxplots (with points superimposed) showing the root mean square prediction errors from each model for the low noise scenarios. The median values are shown at the top of each boxplot for clarity.}
    \label{fig:RMSPElow}
\end{figure}

\begin{figure}
    \centering
   \includegraphics[width=0.9\linewidth]{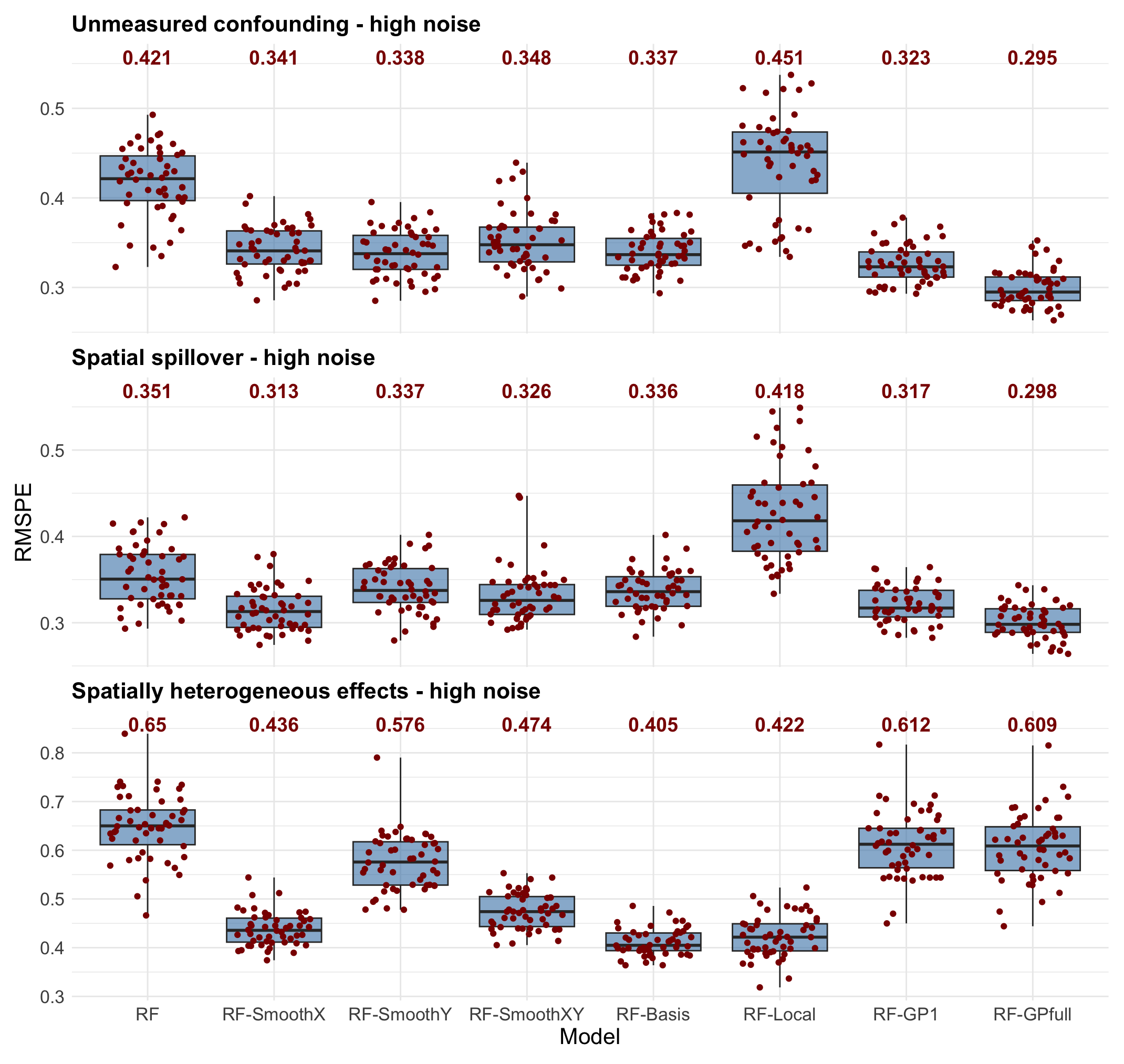}
    \caption{Boxplots (with points superimposed) showing the root mean square prediction errors from each model for the high noise scenarios. The median values are shown at the top of each boxplot for clarity.}
    \label{fig:RMSPEhigh}
\end{figure}

The performance of each model's point prediction is measured by bias and root mean square prediction error (RMSPE), with the former  assessing whether each model over or under predicts on average, while the latter measures the precision in the point prediction. Mathematical definitions of these quantities are given in Section S3.1 of the supporting material. All models produce essentially unbiased predictions in all 6 scenarios, and the full results are not presented for brevity. In summary, the mean biases over the 50 simulated data sets for each model in each scenario range between -0.021 and 0.009, which compares to an average data value of around 10. Boxplots of the RMSPEs across the 50 simulated data sets for each model and scenario are presented in Figures \ref{fig:RMSPElow} (low noise scenarios) and \ref{fig:RMSPEhigh} (high noise scenarios), where the data set specific values are superimposed as points for clarity. In each figure the 3 panels relate to the different causes of spatial autocorrelation. In each case the median values over the 50 simulated data sets are presented above each boxplot for clarity.

The figure shows a number of interesting results, the first of which is that on average across the six scenarios the standard non-spatial random forest \texttt{RF} provides the worst predictive performance with the highest RMSPEs. This is unsurprising, and is likely to be because it does not utilise the spatial autocorrelation in the data when making predictions. The local spatial fitting model \texttt{RF-Local} naturally performs very well when the data exhibit spatially heterogeneous covariate-response relationships, having close to the best RMSPE in this setting. However, it performs very poorly in the other scenarios, which is because the flexibility of local fitting is not required and as its local predictions are based on a relatively small number of data points its prediction error increases. This is especially true in the high-noise scenarios in Figure \ref{fig:RMSPEhigh}. The Gaussian process-based models \texttt{RF-GP1} and \texttt{RF-GPfull} are also inconsistent, performing the best if the cause of the spatial autocorrelation is unmeasured confounding or spatial spillover effects, but very poorly if there are spatially heterogeneous covariate-response relationships. In contrast, using spatial basis functions or spatially smoothed features provides relatively consistent results across the six scenarios, with relative performances that are: (i) always better than a non-spatial random forest; and (ii) relatively close to the best performing model.  Of these, the spatial basis function approach (\texttt{RF-Basis}) appears to be the best, with consistently low RMSPE across all 6 of the scenarios considered here.

\subsubsection{Speed of implementation}
Finally, we examine how long each of these models takes to run, which partially quantifies how practical they are for others to implement. The full simulation study was run on compute servers due to their computational convenience, but this would not give researchers a realistic summary of how long it would take to fit each model on their own desktop computer. Therefore, we assess run times by applying each model to five simulated data sets generated under a single scenario, and all results are obtained on an iMac desktop computer with a 3.8 GHz 8-Core Intel Core i7 processor and 32 GB of memory.

Run times in seconds for each model are consistent across the five simulated data sets, with coefficients of variation (standard deviation divided by the mean)  being between 0.013 and 0.054. Unsurprisingly a simple non-spatial random forest (\texttt{RF}) is the fastest to implement taking only 16 seconds on average. The next fastest was the one-step Gaussian process model (\texttt{RF-GP1}) at only 21 seconds, while the three smooth feature models (\texttt{RF-SmoothX}, \texttt{RF-SmoothY} and \texttt{RF-SmoothXY}) ranged between 57 and 94 seconds on average. The basis function (\texttt{RF-Basis}) and full Gaussian process models (\texttt{RF-GPfull}) took 265 and 202 seconds on average respectively, which while being around 3 time longer than the smooth feature models are still only around 4 minutes of computing time. Finally, the outlier in fitting time is the local fitting model (\texttt{RF-Local}) at 4,167 seconds on average, which is around 1 hour and 10 minutes. As a result, it is the only model of those considered that doesn't run in near real-time, which is due to it fitting a separate local random forest for each data point.

\section{Results from the Blantyre SCALE study}
Section 5.1 presents a validation study quantifying each model's out-of-sample predictive ability, while Section 5.2 investigates the relative importance of the  covariates. Finally, Section 5.3 uses the best performing model to predict PM$_{2.5}$ concentrations for the $N_2=3,950$ households without PM$_{2.5}$ measurements.

\subsection{Validation study}

\subsubsection{Study design}
The response variable is the natural log of the median PM$_{2.5}$ concentration at each household for the reasons discussed in Section 2, while all predictive quality metrics are applied to predicted concentrations  back-transformed to the original scale. The set of covariates utilised are those listed in Table 1 in the supporting material, and all continuous covariates were standardised to have a mean of zero and a standard deviation of one. The same set of random forest frameworks compared in the simulation study are compared in this study, specifically \texttt{RF}, \texttt{RF-SmoothX}, \texttt{RF-SmoothY}, \texttt{RF-SmoothXY}, \texttt{RF-Basis}, \texttt{RF-Local}, \texttt{RF-GP1} and  \texttt{RF-GPfull}. Predictive performance is quantified by splitting the $N_1=3,223$ observed data points  into an 80\% training set (2,578 households) and a 20\% test set (645 households), and this random out-of-sample prediction experiment is repeated five times to ensure the results are not biased by a single choice of training-test split. Each random forest model is fitted with 2,000 trees, and the tuning parameters are optimised as discussed in Section 3. 

After tuning, each model is refitted to the full training set and used to make out-of-sample predictions with 95\% prediction intervals for observations in the test set. As the predictions are likely to be approximately normally distributed on the log scale, the following well known log-normal result is applied to the point predictions to ensure they are approximately unbiased: if $\mathbf{X}\sim\mbox{N}(\mu, \sigma^2)$, then $\mathbf{E}[\exp(X)]=\exp(\mu + \sigma^2/2)$, where $\sigma^2$ is approximated using the out-of-bag prediction mean square error. The predictive performance of each model is assessed using the same metrics as in the simulation experiment, namely bias, RMSPE, CP and MIW. 

\subsubsection{Results of the predictive assessment}
Initially, we investigated the impact of using weighted (based on the MAD) or unweighted model fitting for a standard random forest, because the median PM$_{2.5}$ concentrations measured at each household exhibit different levels of within-household variability. The results of this study are displayed in Section S4.1 of the supporting material, and show that using standard unweighted fitting  produces slightly better results to the best weighted alternative. Thus, the additional complexity induced by weighted fitting when making predictions for households with no data  (i.e., no MAD value is available) seems unnecessary, and hence unweighted fitting is used hereafter. We then compared the predictive performance of a standard random forest against alternative prediction methodologies, to assess whether the random forest algorithm is the most appropriate framework for predicting PM$_{2.5}$ concentrations in the SCALE study. The results of this additional study are presented in Section S4.2 in the supporting material, and show that random forests are the optimal prediction algorithm in this context. 

The results of the main validation study are presented in Table \ref{table_smoothed}, which presents results for each training-test split as well as the mean over all five splits. The table shows that while all models produce close to unbiased predictions there is a slight tendency for under-prediction on the original scale, with mean biases ranging between -1.27 and -0.56. The standard non-spatial RF exhibits the worst performance in terms of point prediction, because it has the highest RMSPE of 15.33$\mu gm^{-3}$ compared to the next highest of 14.99$\mu gm^{-3}$. It also exhibits the widest 95\% prediction intervals, despite having largely comparable coverage probabilities with the other models that range between 0.94 and 0.97. This illustrates that extending the RF algorithm to account for spatial autocorrelation leads to consistent improvements in predictive performance, regardless of how that spatial information is incorporated into the algorithm. 

The table also shows that for the SCALE study the geographical random forest algorithm (\texttt{RF-Local}) exhibits the best predictive performance, having the smallest RMSPE by around 0.3$\mu gm^{-3}$ compared to the next best model (\texttt{RF-Basis}). It also has coverage probabilities that are closest to 0.95, while having relatively narrow 95\% prediction intervals compared to its competitors. However, as the simulation study has evidenced it is by far the most computationally intensive to implement, making the basis function approach (\texttt{RF-Basis}) a plausible competitor as it only has slightly worse predictive performance while being much faster to fit. The remaining approaches that use spatially smoothed features or a Gaussian process to account for the spatial autocorrelation in the data perform similarly, with little to choose between them in terms of predictive performance. Finally, the table shows that PM$_{2.5}$ concentrations can be predicted with relatively good accuracy, because the best average RMSPE of 14.03$\mu gm^{-3}$ is relatively small compared to the range of data values between 0.9$\mu gm^{-3}$ and 297.6$\mu gm^{-3}$.

\subsection{Quantifying covariate importance}
The relative importance of each covariate in explaining / predicting the natural log of PM$_{2.5}$ concentrations is measured by a variable importance plot (VIP, see \citealp{boehmke2020} for a description). For a single decision tree the permutation-based importance of each covariate is calculated by first making out-of-bag predictions for the subset of data points not used to fit that particular tree, before computing their root mean square prediction error. Then the values for a single covariate are randomly permuted and the OOB RMSPE is re-computed. The importance of that covariate is measured by the reduction in predictive accuracy resulting from this random permutation step, and covariates associated with the largest increase in OOB RMSPE are considered the most important. This single variable and single tree importance is then averaged over all the decision trees in the forest for each covariate, before scaling these covariate specific importances to the interval $[0,100]$ for ease of presentation. 

\begin{figure}
    \centering
    \includegraphics[width=0.9\linewidth]{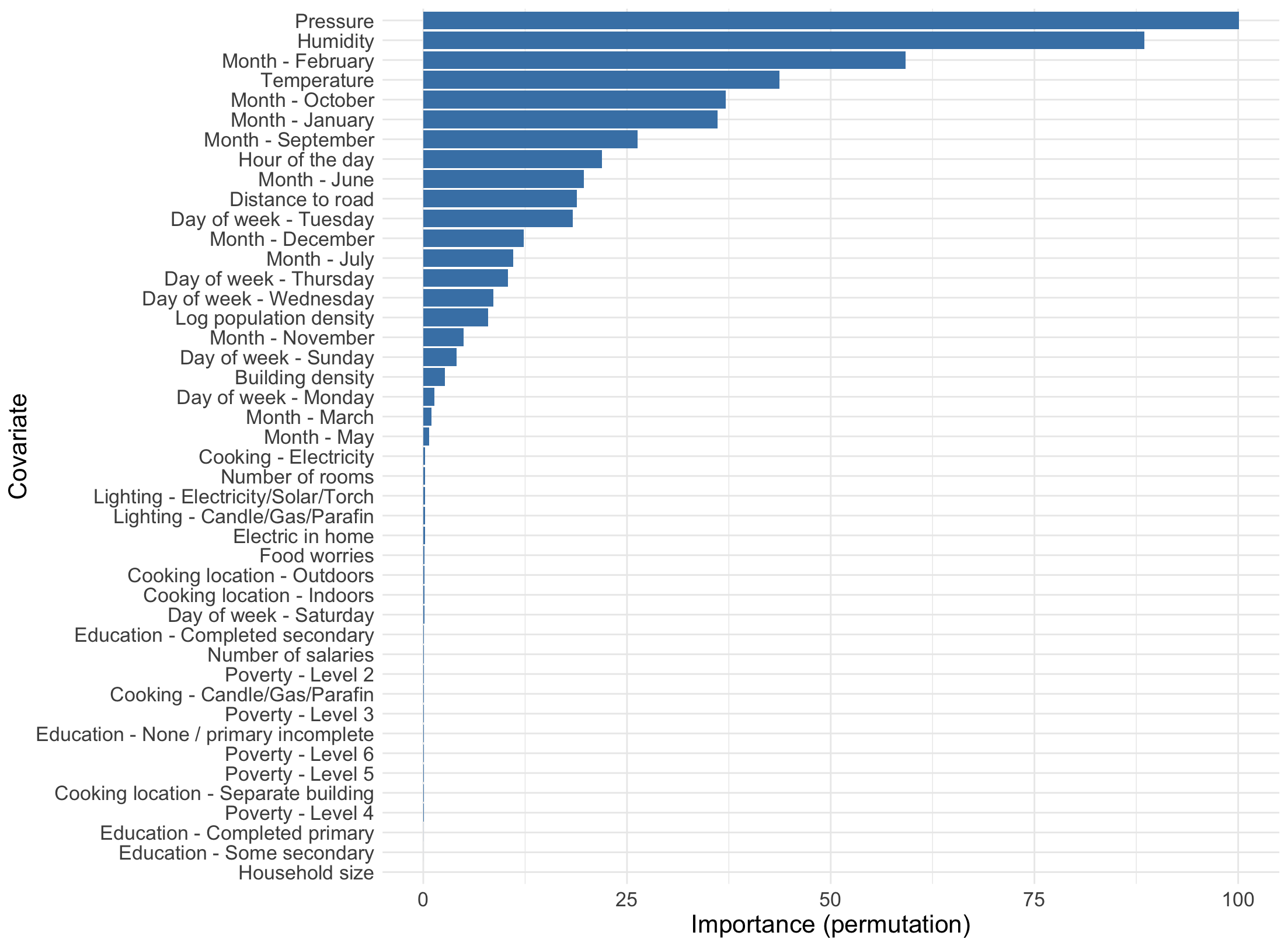}
    \caption{Variable importance plot showing the relative importance of each covariate in a global random forest model.}
    \label{fig:vip}
\end{figure}

The VIP from a global random forest model is presented in Figure \ref{fig:vip}, and shows that the meteorology and month variables are the most important covariates, with the three meteorology variables placing 1$st$ (pressure), 2$nd$ (humidity) and 4$th$ (temperature) most important,  while the 10 monthly indicators (August is excluded as its the baseline level) range between 3$rd$ and 22$nd$ most important. The other notably important variables include the hour of the day (8$th$), the distance to the nearest road (10$th$), the day of the week (highest level is 11$th$), and log population density (16$th$). In contrast, the measures of poverty and household utilities (e.g., main cooking source) appear to be much less important, being in the bottom half of variable importances in all cases. 

\subsection{PM$_{2.5}$ predictions for households with no measurements}   
The validation study  presented in Section 5.1 shows that a geographical random forest (\texttt{RF-Local}) provides the best out-of-sample predictive performance, so here we use this model to predict the concentrations for the $N_2=3,950$ households in the SCALE study where PM$_{2.5}$ concentrations were not measured. 

The distributions of the predictions compared with the observed concentrations are displayed in Section S4.3 of the supporting material via histograms, while maps are displayed in Figure \ref{fig:obspred} below. The figure shows that the spatial pattern in the predictions appears to be consistent with that of the observed data, with the highest concentrations occurring in the south and east of the region while the lowest concentrations appear in the north and central areas. Additionally, the predictions show a smoothing out of the extreme observed values as expected, such as the apparent high outlying (darkest on the map) values in the south of the region.

\begin{figure}[p]
    \centering
    \begin{subfigure}{\textwidth}
        \centering
        \includegraphics[width=0.70\linewidth]{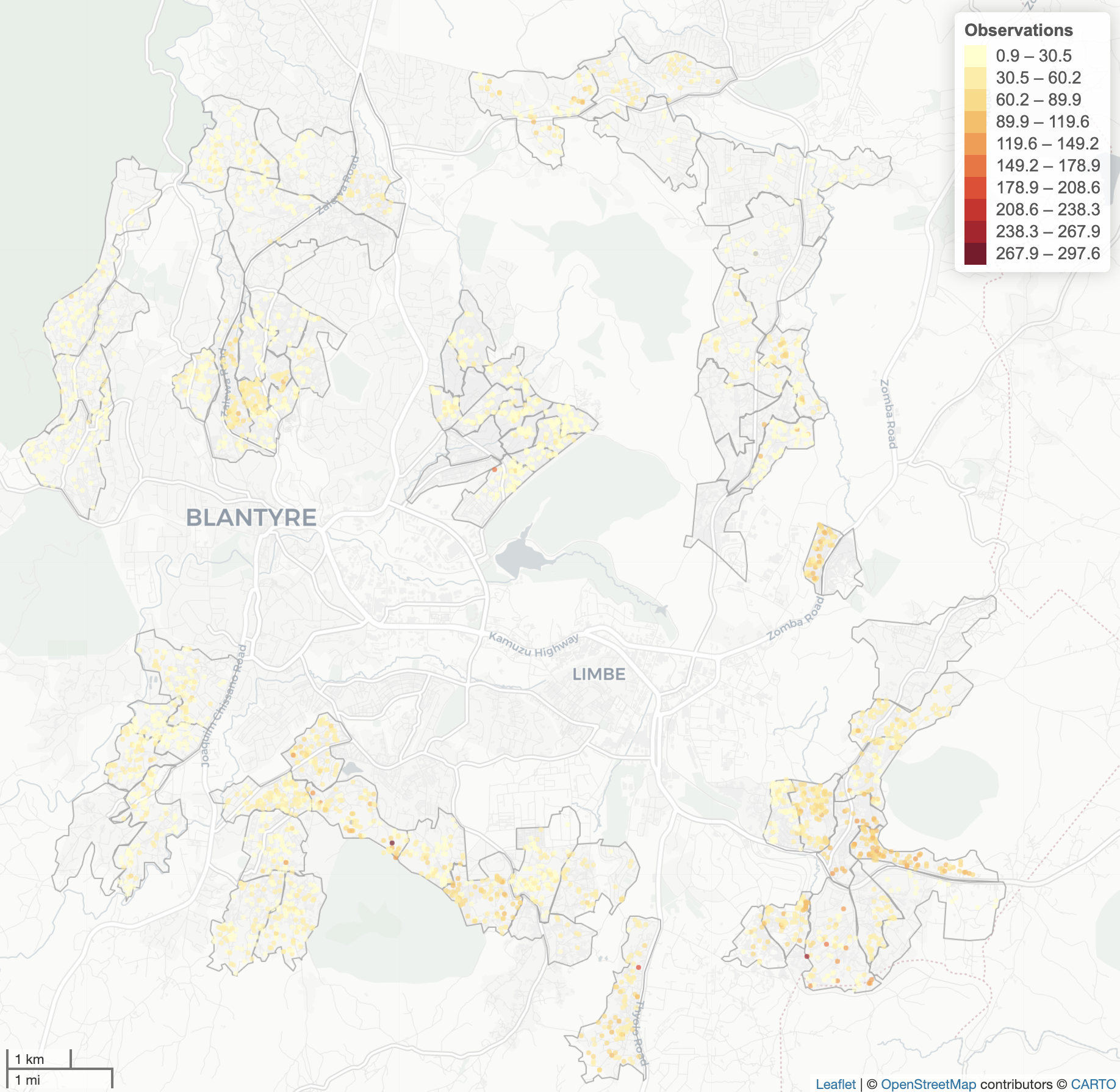}
        \caption{Observed concentrations.}
    \end{subfigure}
    \vfill 
    \begin{subfigure}{\textwidth}
        \centering
        \includegraphics[width=0.70\linewidth]{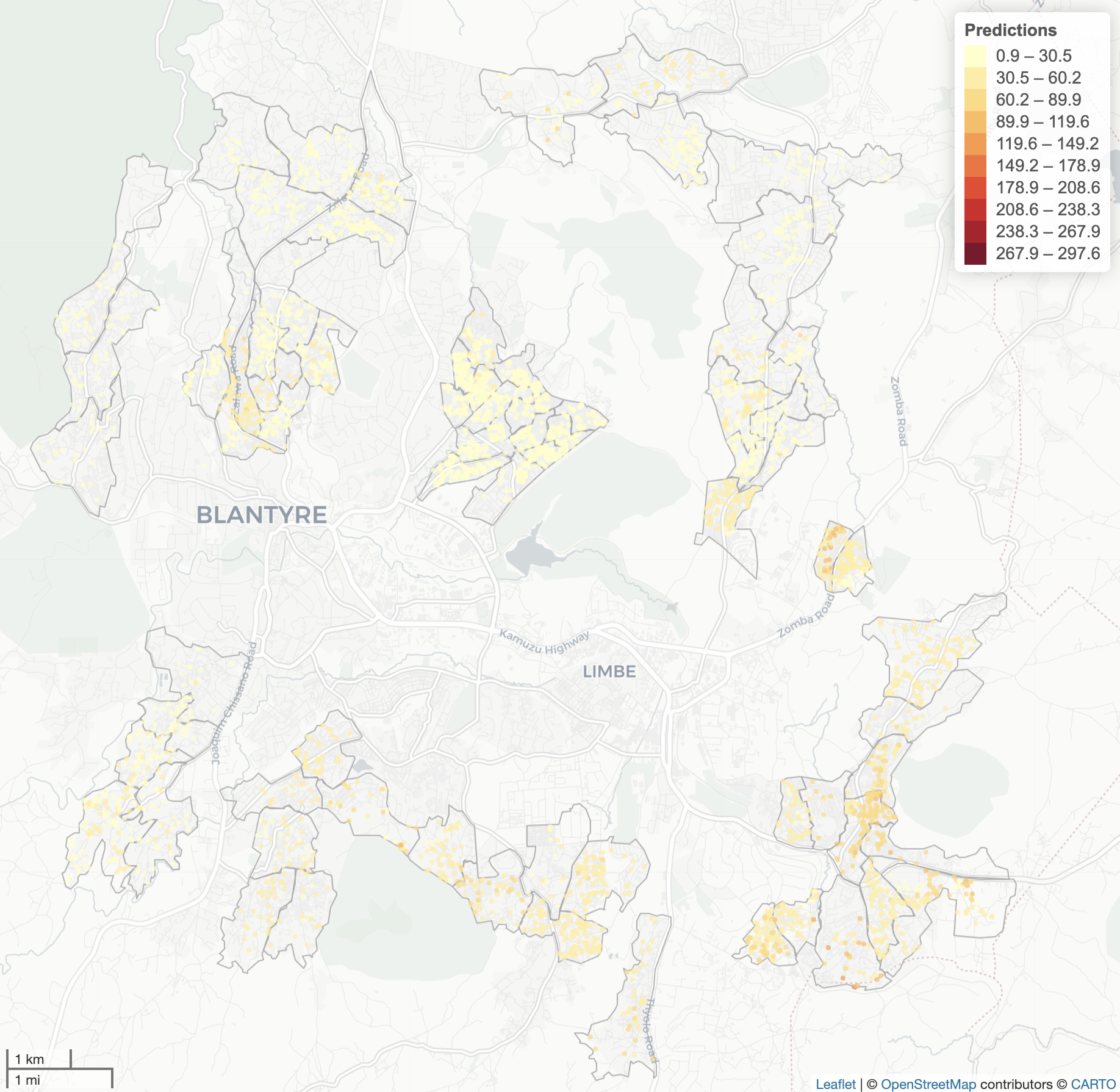}
        \caption{Predicted concentrations.}
    \end{subfigure}

    \caption{Maps of the observed (top) and predicted (bottom) PM$_{2.5}$ concentrations in $\mu gm^{-3}$.}
    \label{fig:obspred}
\end{figure}

\section{Discussion}

This paper has presented the first in-depth performance comparison of a range of spatially adjusted random forest algorithms for predicting spatially autocorrelated geostatistical data, which extends the recent narrative taxonomy presented by \cite{Patelli2024}. Our performance comparison is based on both a comprehensive simulation experiment and a real data case study, the latter focusing on predicting particulate matter concentrations in a low and middle income country setting. Collectively, our studies yield the following key findings, which will be especially  relevant to researchers tasked with predicting spatially resolved environmental processes. 
    
The first is that almost any sensible adaptation of the standard random forest algorithm to account for the residual spatial autocorrelation ubiquitous in spatial data improves predictive performance, in terms of both point prediction and the precision of its uncertainty quantification. The second is that no single approach is optimal across the range of spatial autocorrelation mechanisms considered here, suggesting there is not a one-size-fits-all best approach to allowing for spatial autocorrelation in a random forest prediction algorithm. This implies that the optimal prediction model for a given data set can only be determined by comparing multiple prediction models on the data under study, for example by adopting an out-of-sample prediction experiment similar to that presented in section 5. However, such a comparative approach is clearly time consuming, so for researchers wishing to utilise a single prediction algorithm then a \emph{safe bet} is to use the basis function approach, because it is relatively fast to implement and it exhibited close to the smallest prediction error across a wide range of the data sets considered here.   

The spatial sampling design utilised in the SCALE study was naturally done for convenience of the fieldworkers and to minimise costs, but was sub-optimal from a modelling perspective because each site was only sampled once and nearby sites were typically sampled on the same or days close together. Thus, as there was no replication of the spatial surface over time it was not possible to disentangle the effects of space and time, as either one could be driving the variation in the observed data. The variable importance plot summarising the relative importance of the covariates shows that meteorology (varying temporally) and time indicator variables are by far the most dominant covariates. In contrast, household specific characteristics such as the cooking fuel used, level of poverty, etc were much less important determinants of  PM$_{2.5}$ concentrations. This suggests that either time is  the most important factor influencing PM$_{2.5}$ concentrations, or that because time is acting as a proxy for space in the data collection design, it is space that is the key determining factor in the concentrations. All of this suggests that in future studies of this type, repeated monitoring of the households at different points in a year would be hugely beneficial for accurately predicting spatio-temporal air pollution fields.  

The use of spatially adapted machine learning algorithms for spatial prediction is one of the current cutting edge research themes in spatial analytics, and as a result there are numerous avenues of exciting future research directions. The first is how to extend spatially adapted ML algorithms for multivariate spatio-temporal data, such as predicting the concentrations of multiple correlated pollutants in both space and into the future. The challenges inherent in this endeavor include how to accurately capture the spatio-temporal and cross correlation structures in the data, as well as the computational issues arising from large data sizes and increased numbers of spatial and temporal tuning parameters. A second open question is, is there a best ML algorithm to adapt for use with spatial data,  or like the choice of spatial adaptation method, does the choice between a random forest, neural network, gradient boosting machine, etc, depend on the data under study. Finally, a future epidemiological project to be undertaken as part of the SCALE study is to use the PM$_{2.5}$ predictions produced here to examine whether close proximity to higher pollution concentrations increases the risk of tuberculous. 

\backmatter

\section*{Declarations}

\begin{description}
    \item[\textbf{Supporting material}] 
    Additional information and supporting material for this article is available online at the journal's website
    
    \item[\textbf{Acknowledgements}]

    \item[\textbf{Ethics}] The SCALE trial was approved by the research ethics committees of the London School of Hygiene \& Tropical Medicine (16228) and the Kamuzu University of Health Sciences (P.12/18/2556). We obtained written informed consent to participate from all individuals interviewed. Where participants were illiterate, independently-witnessed thumbprint consent was recorded. 
    \item[\textbf{Funding}] HRS was funded by the UK Medical Research Council through the Medical Research Council Doctoral Training Partnership Programme at Liverpool School of Tropical Medicine (MR/N013514/1). PM was funded by Wellcome (304666/Z/23/Z) and an NIHR Global Health Research Professorship (NIHR304311). The views expressed are those of the author(s) and not necessarily those of the NIHR or the Department of Health and Social Care.
    
    \item[\textbf{Open Access}] For the purpose of open access, the author has applied a Creative Commons Attribution (CC BY) licence to any Author Accepted Manuscript version arising from this submission.

    \item[\textbf{Competing interests}] None.

    \item[\textbf{Data and code availability}] The data are the property of the SCALE study and cannot be made publicly available. However, a summary of the code and simulated data available with this paper is given in Section S2.2 of the supporting material.
\end{description}

\bibliography{malawi_main.bib}

\newpage
\begin{table}
    \centering
    \caption{Comparison of the out-of-sample predictive abilities of a non-spatial random forest and a number of spatially adapted alternatives. The table presents bias, root mean square prediction error (RMSPE), coverage probability (CP) and mean interval width (MIW) for each of the five training and test splits as well as their mean.}
    \label{table_smoothed}
    \vspace{5pt}
    \begin{tabular}{lrrrrrr}
        \toprule
        \multirow{2}{*}{\textbf{Model}} & \multicolumn{6}{c}{\textbf{Split}} \\
        \cmidrule(lr){2-7}
        & \textbf{1} & \textbf{2} & \textbf{3} & \textbf{4} & \textbf{5} & \textbf{Mean} \\
        \midrule
        \addlinespace
        \textbf{Bias} \\
        \texttt{RF}         &-0.58 & -0.90 & -1.89 & -1.05 & -0.82 & -1.05 \\
        \texttt{RF-SmoothX} &-1.49 & -0.84 & -1.72 & -1.10 & -1.20 & -1.27 \\  
        \texttt{RF-SmoothY} &-0.12 & -1.29 & -2.13 & -1.42 & -0.56 & -1.10 \\ 
        \texttt{RF-SmoothXY} &-0.80 & -1.35 & -2.08 & -1.24 & -0.86 & -1.27 \\  
        \texttt{RF-Basis} &-0.25 & -0.63 & -1.19 & -0.63 & -0.53 & -0.65 \\ 
        \texttt{RF-Local} &0.06 &-0.37 &-0.38 &-0.06  &0.21   &-0.11 \\
        \texttt{RF-GP1} &-0.02 & -0.47 & -1.31 & -0.51 & -0.50 & -0.56 \\  
        \texttt{RF-GPfull} &-0.10 & -0.51 & -1.40 & -0.51 & -0.71 & -0.65 \\ 
        \midrule
        \addlinespace
        \textbf{RMSPE} \\
        \texttt{RF}          &12.27 & 15.50 & 18.07 & 15.57 & 15.26 & 15.33 \\
        \texttt{RF-SmoothX} &  12.19 & 14.97 & 17.61 & 14.73 & 14.44 & 14.79 \\  
        \texttt{RF-SmoothY} &11.90 & 14.33 & 17.89 & 15.53 & 14.40 & 14.81 \\ 
        \texttt{RF-SmoothXY} &12.19 & 15.07 & 17.99 & 14.73 & 14.32 & 14.86 \\  
        \texttt{RF-Basis} &11.66 & 14.07 & 17.32 & 14.43 & 14.17 & 14.33 \\
        \texttt{RF-Local} &10.98 &14.06 &17.05 &13.94 &14.11   &14.03 \\
        \texttt{RF-GP1} &  11.96 & 15.12 & 17.64 & 15.43 & 14.81 & 14.99 \\ 
        \texttt{RF-GPfull} &11.85 & 15.02 & 17.93 & 15.29 & 14.81 & 14.98 \\ 
        \midrule
        \addlinespace
        \textbf{CP} \\
        \texttt{RF} &0.97  &0.97  &0.95  &0.97  &0.97  &0.97  \\
        \texttt{RF-SmoothX} &0.98 & 0.97 & 0.94 & 0.96 & 0.97 & 0.96 \\ 
        \texttt{RF-SmoothY} &0.97 & 0.97 & 0.95 & 0.97 & 0.97 & 0.97 \\ 
        \texttt{RF-SmoothXY} &0.98 & 0.97 & 0.95 & 0.97 & 0.97 & 0.97 \\ 
        \texttt{RF-Basis} &0.97 & 0.96 & 0.94 & 0.96 & 0.96 & 0.96 \\
        \texttt{RF-Local} &0.96  &0.94  &0.94  &0.97  &0.95    &0.95 \\
        \texttt{RF-GP1} &0.95 & 0.93 & 0.94 & 0.94 & 0.94 & 0.94 \\ 
        \texttt{RF-GPfull} &0.95 & 0.93 & 0.94 & 0.94 & 0.94 & 0.94 \\ 
        \midrule
        \addlinespace
        \textbf{MIW} \\
        \texttt{RF}    &63.45  &58.52  &57.39  &59.18  &58.15  &59.34  \\
        \texttt{RF-SmoothX} &  62.87 & 58.69 & 54.91 & 57.42 & 57.56 & 58.29 \\ 
        \texttt{RF-SmoothY} &61.77 & 54.09 & 53.70 & 59.47 & 55.66 & 56.94 \\ 
        \texttt{RF-SmoothXY} &64.36 & 57.51 & 53.87 & 59.20 & 56.24 & 58.24 \\ 
        \texttt{RF-Basis} &58.20 & 51.28 & 51.42 & 54.07 & 53.49 & 53.69 \\
        \texttt{RF-Local} &55.97 &52.28 &53.45 &54.41 &53.88   &54.00 \\
        \texttt{RF-GP1} &54.34 & 50.43 & 50.52 & 52.55 & 50.91 & 51.75 \\ 
        \texttt{RF-GPfull} &53.94 & 50.23 & 50.13 & 52.05 & 50.45 & 51.36 \\ 
        \bottomrule
    \end{tabular}
\end{table}

\end{document}